\documentclass{elsarticle}
\usepackage{graphicx}
\usepackage{amsmath}
\usepackage{amsfonts}
\usepackage{amssymb}

\newcommand{\kB}{{k_{\mathrm B}}}
\newcommand{\Tmin}{{T_{\mathrm{min}}}}

\begin{document}
\journal{Phys. Lett. A}

\begin{frontmatter}
\title{Proposed lower bound for the shear viscosity to entropy density ratio in
some dense liquids}

\author[CTPhys,CNISM,INFN]{G. G. N. Angilella\corref{corr}}
\author[RUCA,ICTP,Oxford]{N. H. March}
\author[SSC,CTPhys,INFN]{F. M. D. Pellegrino}
\author[CTPhys,CNISM]{R. Pucci}

\address[CTPhys]{Dipartimento di Fisica e Astronomia, Universit\`a di
Catania,\\ Via S. Sofia, 64, I-95123 Catania, Italy}
\address[CNISM]{CNISM, UdR Catania, Italy}
\address[INFN]{INFN, Sez. Catania, Italy}
\cortext[corr]{Corresponding author.}
\address[RUCA]{Department of Physics, University of Antwerp,\\
Groenenborgerlaan 171, B-2020 Antwerp, Belgium}
\address[ICTP]{The Abdus Salam International Centre for Theoretical Physics,\\
Strada Costiera, 11, Miramare, Trieste, Italy}
\address[Oxford]{Oxford University, Oxford, UK}
\address[SSC]{Scuola Superiore di Catania, Universit\`a di Catania,\\
Via S. Nullo, 5/i, I-95123 Catania, Italy}

\begin{abstract}
Starting from relativistic quantum field theories, Kovtun \emph{et al.} (2005)
have quite recently proposed a lower bound $\eta/s\geq \hbar /(4 \pi \kB)$,
where $\eta$ is the shear viscosity and $s$ the volume density of entropy for
dense liquids. If their proposal can eventually be proved, then this would
provide key theoretical underpinning to earlier semiempirical proposals on the
relation between a transport coefficient $\eta$ and a thermodynamic quantity
$s$. Here, we examine largely experimental data on some dense liquids, the
insulators nitrogen, water, and ammonia, plus the alkali metals, where the shear
viscosity $\eta(T)$ for the four heaviest alkalis is known to scale onto an
`almost universal' curve, following the work of Tankeshwar and March a decade
ago. So far, all known results for both insulating and metallic dense liquids
correctly exceed the lower bound prediction of Kovtun \emph{et al.}
\end{abstract}
\end{frontmatter}

\section{Background and outline}

Going back, at very least, to the early semiempirical proposal of Rosenfeld
\cite{Rosenfeld:77}, quite an active field has grown up relating shear viscosity
$\eta$ to the volume density of entropy $s$ in dense liquids
\cite{Dzugotov:96,Hoyt:00,Kaur:05}. Some motivation for returning to this
general area, however, has come from unexpected quarters, namely relativistic
quantum field string theory methods. Thus, Kovtun \emph{et al.}
\cite{Kovtun:05}, starting from such an \emph{ab initio} viewpoint, have
proposed that the ratio $\eta/s$ has a lower bound given by $\hbar/(4\pi\kB)
\approx 6.08\cdot 10^{-13}$~K$\cdot$s. Here, $h$ and $\kB$ denote Planck's and
Boltzmann's constants, respectively. While the lower bound proposal is not yet
proven, their arguments are highly plausible (see Section~\ref{sec:summary} for a
brief summary).

The outline of the present paper is then as follows. After a brief review of the
hard sphere model in Section~\ref{sec:hs}, with specific emphasis to its results
for the shear viscosity and entropy density in dense fluids, in
Section~\ref{sec:liquids}, we examine some scaling properties of the three dense
insulating liquids He, N$_2$, and H$_2$O considered in \cite{Kovtun:05}, as well
as NH$_3$. This is then followed in Section~\ref{sec:transport} by work on dense
liquid metals \cite{March:90a}. Some six liquid transition metals treated
computationally by means of the embedded atom model are then considered, with
the alkalis playing the dominant role because of already known scaling
properties of the transport coefficient $\eta(T)$ for the four heaviest alkali
metals, Na through Cs \cite{Tankeshwar:98}, these results constituting
Section~\ref{sec:alkali}.  A discussion follows in Section~\ref{sec:summary},
together with proposals for further work that should prove fruitful.

\section{Hard sphere model of shear viscosity $\eta$ and entropy density
\lowercase{$s$} in a dense fluid used in phenomenological context}
\label{sec:hs}

We begin with what we believe is the simplest model of a dense fluid,
\emph{viz.} that composed of hard spheres (HS). In early work, Collins and Raffel
\cite{Collins:54} and Longuet-Higgins and Pople \cite{Longuet-Higgins:56}
stressed that in a liquid of rigid molecules the singular nature of the
interaction allows a finite flux of momentum and energy even when the liquid
radial distribution function is momentarily isotropic, as it is in equilibrium.

The results obtained by Longuet-Higgins and Pople \cite{Longuet-Higgins:56} can
be displayed in terms of the deviation of the equation of state from the ideal
gas form as
\begin{equation}
\eta_{\mathrm{HS}} = \frac{2}{5} \rho \sigma \left( \frac{M \kB T}{\pi}
\right)^{1/2} \left[ \frac{P}{\rho \kB T} -1 \right] ,
\label{eq:LHP}
\end{equation}
where $\sigma$ denotes the hard sphere diameter, while $M$ is the particle mass.
The factor in square brackets in Eq.~(\ref{eq:LHP}) embodies the probability of
finding two spheres at contact, which in turn is given by the contact value of
the pair distribution function $g(r)$.

The Percus-Yevick theory \cite{Percus:58,Thiele:63,Wertheim:63} of the hard
sphere fluid leads to a number of analytic, though of course approximate,
results, which will be used below in exemplifying the hard sphere dense fluid
predictions pertaining to the Kovtun \emph{et al.} \cite{Kovtun:05} lower bound
on the ratio shear viscosity $\eta$ to volume density of entropy $s$.

The so-called excess entropy and other thermodynamic properties of the density
hard sphere fluid can be usually derived with adequate accuracy from the
Carnahan-Starling \cite{Carnahan:69} equation of state. In terms of the packing
fraction $p_f$, this reads
\begin{equation}
\frac{P}{\rho_N \kB T} = \frac{1+ p_f + p_f^2 -p_f^3}{(1-p_f )^3} ,
\label{eq:Carnahan}
\end{equation}
where $p_f = \pi \rho_N \sigma^3 /6$, with $\rho_N = N/V$ being the number density.
Eq.~(\ref{eq:Carnahan}) can be generalized to embrace hard fluids of
non-spherical molecules as \cite{Kolafa:95}
\begin{equation}
\frac{P}{\rho_N \kB T} = \frac{1+ (3\alpha-2)p_f + (\alpha^2 +\alpha-1) p_f^2
-\alpha(5\alpha-4)p_f^3}{(1-p_f )^3} ,
\label{eq:tetrahedra}
\end{equation}
with $\alpha=2.2346$ for tetrahedral molecules, and $\alpha=1$ for spherical
molecules, in which case it reduces to Eq.~(\ref{eq:Carnahan}).
Hence, by specifying the pressure $P$ as in the examples considered by Kovtun
\emph{et al.} \cite{Kovtun:05}, number density $\rho_N$ and temperature $T$ are
related via the packing fraction $p_f$ by Eq.~(\ref{eq:Carnahan}). Insertion of
Eq.~(\ref{eq:Carnahan}) in the right-hand side of Eq.~(\ref{eq:LHP}) then yields
$\eta_{\mathrm{HS}} = \eta_{\mathrm{HS}} (m,\rho,T;\sigma)$.

To construct the Kovtun ratio $\eta/s$ in this model, we next write the excess
entropy $S_E$ in the form
\begin{equation}
\frac{S_{\mathrm{HS}}}{\kB} = \frac{S_E - S_0}{\kB}
= \frac{2}{1-p_f} + \frac{1}{(1-p_f )^2} -3 - \frac{S_0}{\kB} ,
\label{eq:excess}
\end{equation}
where $S_0$ is the entropy for an ideal gas \cite{Landau:80},
\begin{equation}
\frac{S_0}{\kB} = \frac{5}{2} + \log \left[ \frac{1}{\rho_N} \left( \frac{M\kB
T}{2\pi \hbar^2} \right)^{3/2} \right]
= \mathrm{const} + \log T^{3/2} - \log \rho_N ,
\label{eq:S0}
\end{equation}
the latter functional form, including an additive constant, being derived through
thermodynamic considerations only \emph{e.g.} by March and Tosi \cite{March:02}.

\begin{figure}[t]
\centering
\includegraphics[height=0.9\columnwidth,angle=-90]{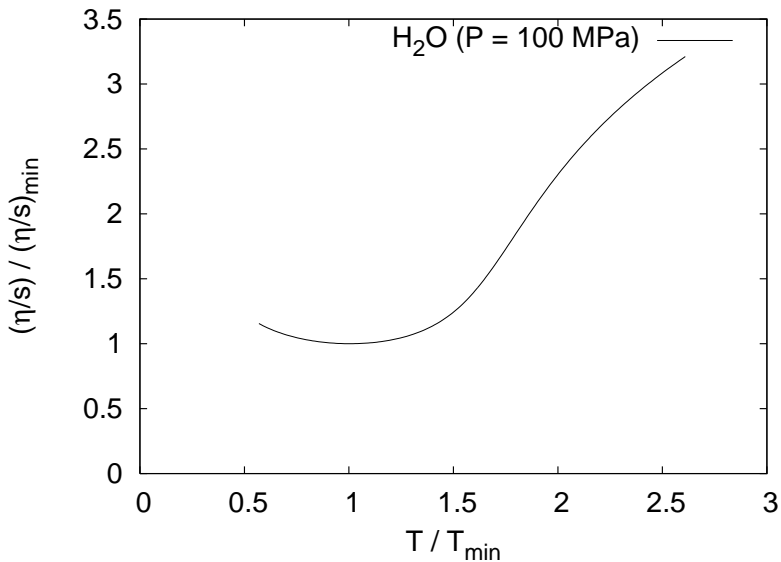}
\caption{Schematic scaled data for ratio $\eta/s$ for $P=100$~MPa, within the
hard sphere model. Data sources referred to in \cite{Kovtun:05}.}
\label{fig:hs}
\end{figure}

The excess entropy $S_E$ in Eq.~(\ref{eq:excess}) is clearly positive in the
range $0<p_f \lesssim 1/2$. Dense fluids, for orientation, freeze when $p_f
\simeq 0.46$. The volume density of entropy $s_{\mathrm{HS}} = \rho
S_{\mathrm{HS}} /\kB$ for the dense hard sphere fluid is hence known from
Eq.~(\ref{eq:excess}). The ratio $\eta/s$ can then be formed from
Eqs.~(\ref{eq:LHP}) and (\ref{eq:excess}), and plotted as a function of $T$,
after eliminating $p_f$ through the hard sphere equation of state
Eq.~(\ref{eq:Carnahan}) choosing the pressure $P=100$~MPa \cite{Kovtun:05}
(Fig.~\ref{fig:hs}). Appendix~\ref{app:D} gives a formal proof relating to
$\eta/s$ under isobaric conditions, that $\eta/s$ is a function of $T$, as in
Fig.~\ref{fig:hs} (see also Fig.~\ref{fig:Kovtun}).

\section{Shape similarity after scaling of available experimental data on dense insulating liquids
helium, nitrogen, water, and ammonia}
\label{sec:liquids}

Though our main focus will be on dense metallic liquids, it seemed natural to
start by examining scaling properties for helium, nitrogen, and water, the
insulating liquids already considered by Kovtun \emph{et al.} \cite{Kovtun:05}
along with their pioneering proposal. In their Fig.~2, all three liquids are
treated in their plot of the ratio $\eta/s$ \emph{vs} absolute
temperature $T$.

\begin{figure}[t]
\centering
\includegraphics[height=0.45\columnwidth,angle=-90]{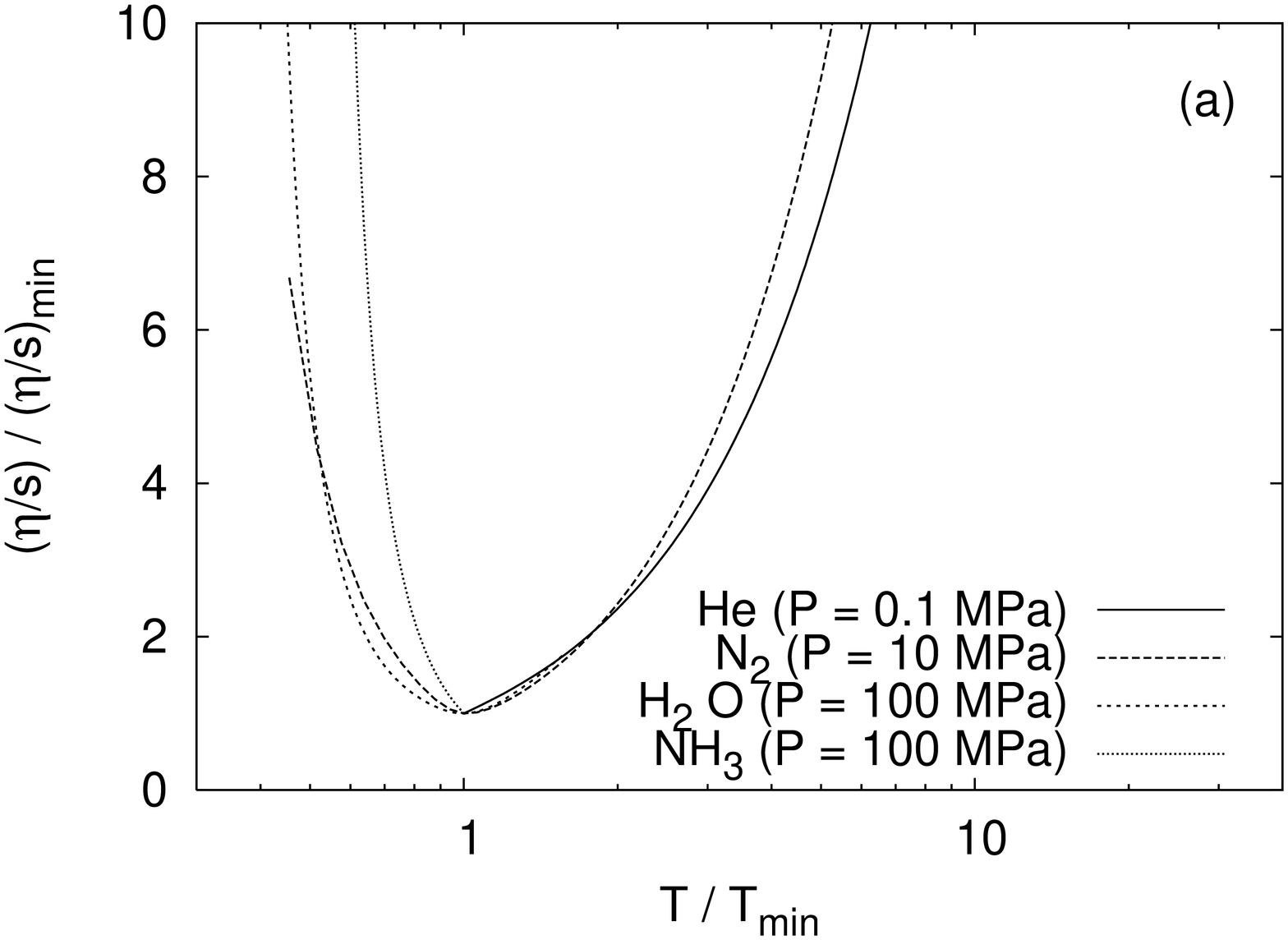}
\includegraphics[height=0.45\columnwidth,angle=-90]{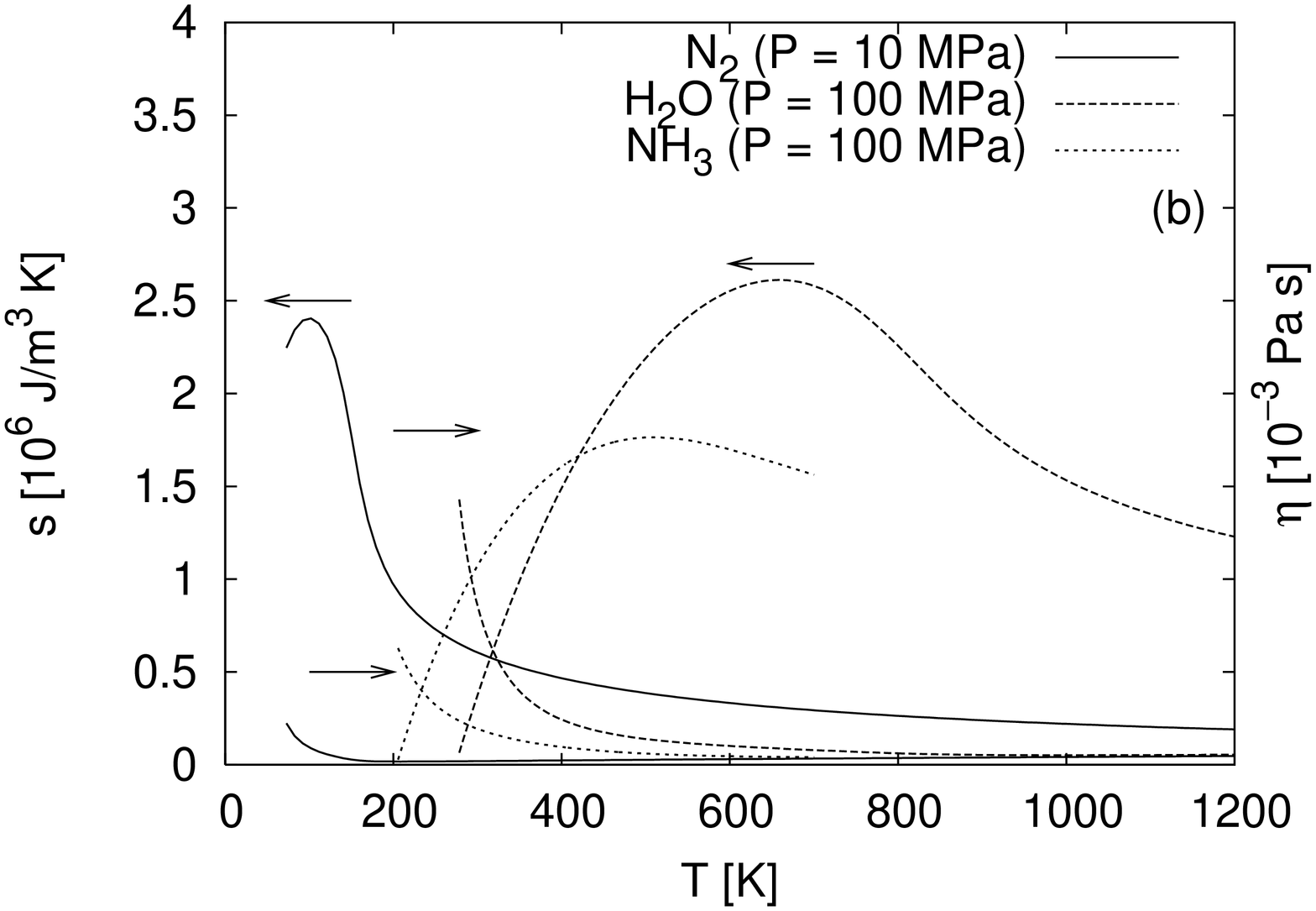}\\
\includegraphics[height=0.45\columnwidth,angle=-90]{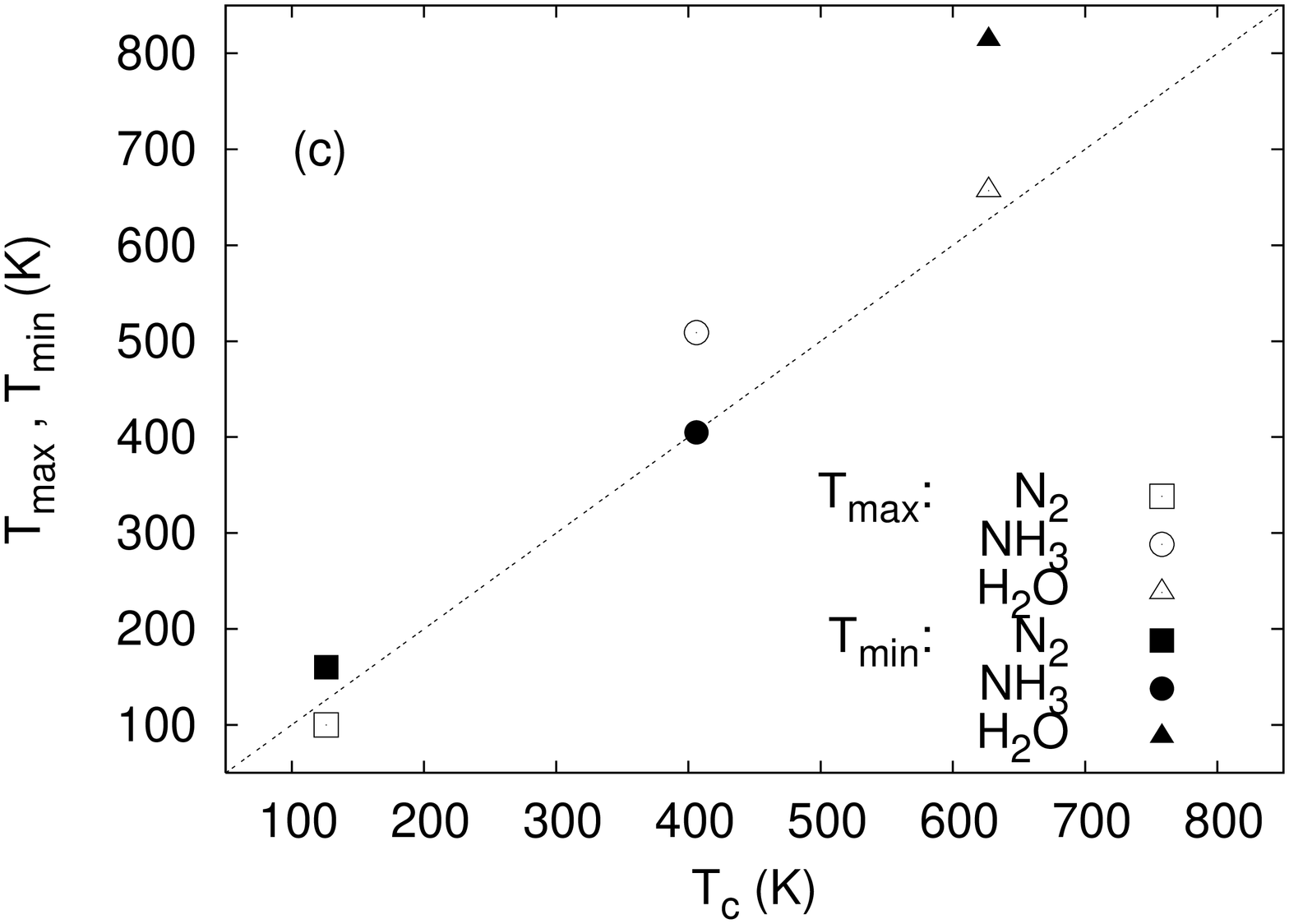}
\caption{(a) Scaled data for ratio $\eta/s$ for insulating dense liquids He,
N$_2$, H$_2$O, and NH$_3$. Data sources referred to in \cite{Kovtun:05}. 
The resulting values (in SI units) for $(\eta/s)_{\mathrm{min}}$ are 
25.07 for He,
24.54 for N$_2$,
45.15 for H$_2$O, and
94.00 for NH$_3$.
(b)
Separate variations with temperature $T$ of volume density of entropy $s$ and of
viscosity $\eta$ (right-hand scale) for nitrogen, water, and ammonia. Data
sources in \cite{Kovtun:05}. (c) Correlation between critical temperatures and 
temperatures $T_{min}$, at which $\eta/s$ is minimum in Fig.~\ref{fig:Kovtun}a
(closed symbols), and temperatures $T_{max}$, at which the entropy density is
maximum in Fig.~\ref{fig:Kovtun}b (open symbols), for nitrogen, water, and
ammonia. The straight line is a guide to the eye.}
\label{fig:Kovtun}
\end{figure}

Because of the lower bound proposal, we have replotted their data in
Fig.~\ref{fig:Kovtun}a. This scales $T$ with $\Tmin$, and also scales
$(\eta/s)_T$ in units of the value at the minimum, $(\eta/s)_\Tmin$. It is of
interest that for these very different liquids, the data collapses somewhat to
exhibit considerable shape similarity. For completeness, as well as to gain
insight as to the source of the minima in Fig.~\ref{fig:Kovtun}a, we show $s(T)$
and $\eta(T)$ separately in Fig.~\ref{fig:Kovtun}b for nitrogen, water, and
ammonia. Over the range of temperatures displayed there, the viscosity is
featureless, and it is the entropy density $s(T)$ that has the characteristic
features. It is noteworthy that the maximum in $s(T)$ for water is near to the
critical temperature $T_c$. This is also true for nitrogen.

Having examined these four dense insulating liquids, we turn to our central
examples, an embedded atom model computer simulation of six liquid transition
metals, and, following that, the dense metallic alkali liquids Li through Cs.

\begin{figure}[t]
\centering
\includegraphics[height=0.9\columnwidth,angle=-90]{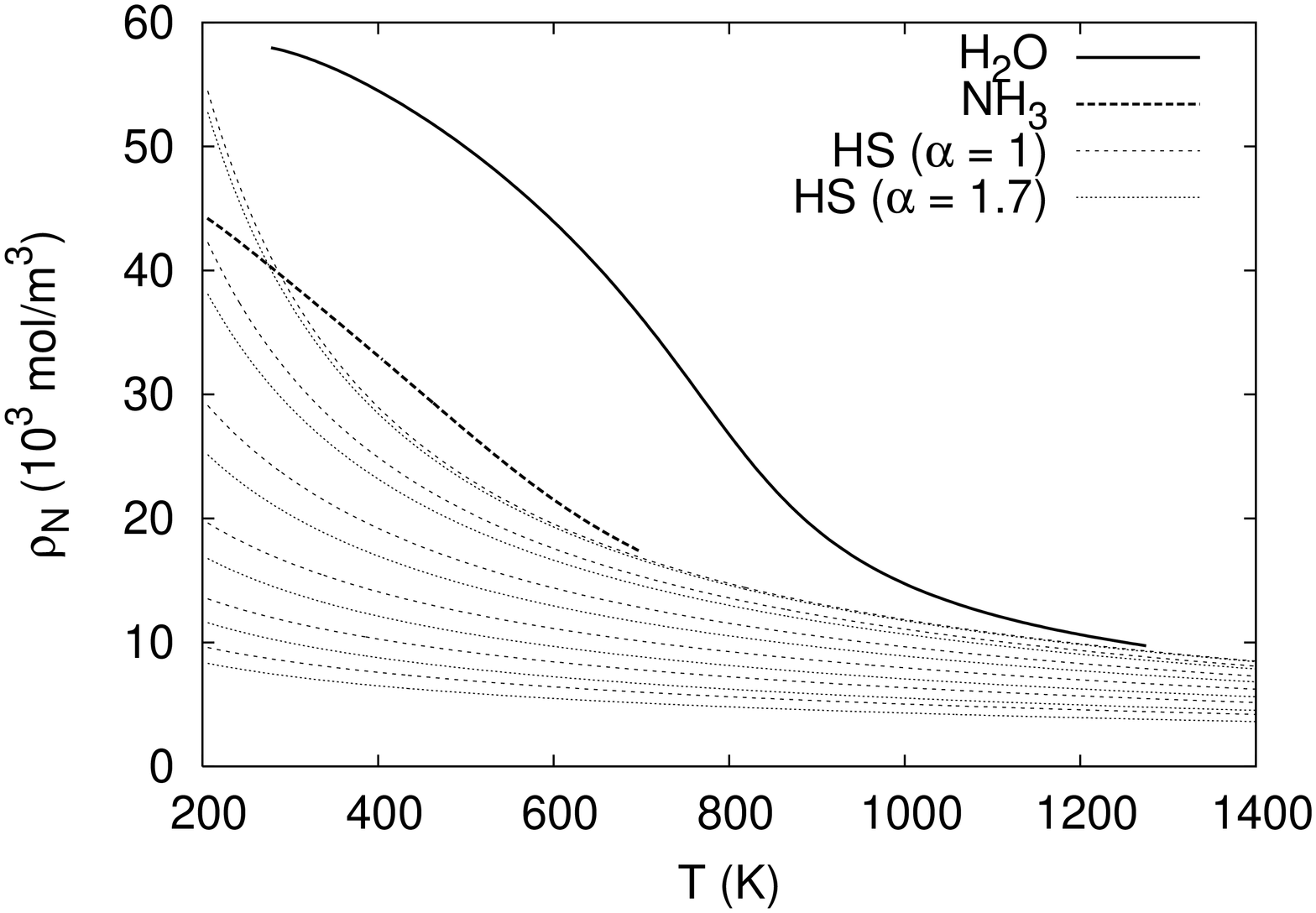}
\caption{Experimental results for the number density $\rho_N$ as a function of
temperature $T$ for H$_2$O and NH$_3$ under isobaric conditions ($P=100$~MPa).
Data sources referred to in \cite{Kovtun:05}. Behavior of modified hard sphere
result in Eq.~(\ref{eq:tetrahedra}) for $\alpha=1$ and 1.7, with $\alpha$ and hard
sphere diameter $\sigma=1-5$~\AA{} treated as parameters, is also shown.}
\label{fig:rhoN}
\end{figure}

\section{Some relevant results concerning transport in liquid metals}
\label{sec:transport}

So far, the specific dense liquids for which we have presented experimental
results are insulators: atomic He and molecular fluids H$_2$O and N$_2$. Though
we do not have such complete experimental data, we feel it of interest to
summarize here some obviously relevant results on some twelve or so liquids
having itinerant electrons, \emph{viz.} liquid transition metals (Cu, \ldots)
and the five liquid alkali metals Li, Na, K, Rb, and Cs.

\subsection{Computer simulation results on some liquid transition metals}

Hoyt \emph{et al.} \cite{Hoyt:00} have carried out important computer
simulation results of liquid transition metals, using as interatomic potential
for their studies those generated by the embedded atom model (EAM). The
transport coefficient they focussed on was the self diffusion $D$. They were
able to refine, for Cu, Ag, Au, Ni, Pd, and Pt, the earlier important study of
Dzugotov \cite{Dzugotov:96}, who approximated the excess entropy via the pair
distribution function $g(r)$. Hoyt \emph{et al.} \cite{Hoyt:00} found a form
like that of the semiempirical proposal of Rosenfeld \cite{Rosenfeld:99}, who
wrote (see also Ref.~\cite{March:02})
\begin{equation}
D \simeq 0.6 \rho_N^{-1/3} \bar{v} \exp (-0.8 s),
\label{eq:Rosenfeld}
\end{equation}
where $s=-S_E /N\kB$, with $S_E$ the excess entropy. Hoyt \emph{et al.}
\cite{Hoyt:00} recovered a form like Eq.~(\ref{eq:Rosenfeld}) for the six
transition metals listed above, but with the factor $\exp(-s)$ to good accuracy.
The scatter of Dzugotov's results \cite{Dzugotov:96} was substantially reduced
by the computer simulation results of Hoyt \emph{et al.}
\cite{Hoyt:00}.

As suitable results for excess entropy are not available to us, we next note two
points: (i) The Stokes-Einstein relation (see Ref.~\cite{March:02}) can be used
to estimate the viscosity of the above transition metals from the Hoyt \emph{et
al.} form for $D$, and (ii) it is of obvious interest to examine the relation of
the above to the predictions of the hard sphere model discussed above. This is
because of the early study of Bernasconi and March \cite{Bernasconi:86} (see
also Ref.~\cite{March:82}). These authors calculated from experiment near
freezing the direct correlation function $c(r)$ at $r=0$, which for hard spheres
the Percus-Yevick theory invoked above predicted to be related to its Fourier
transform, say $\tilde{c} (k=0)$, by \cite{Bhatia:84}
\begin{equation}
\tilde{c} (k=0) = 1 + c(r=0).
\end{equation}

Table~\ref{tab:Bernasconi} shows the values obtained by Bernasconi and March
\cite{Bernasconi:86} for four of the six liquid transition metals considered in
the EAM studies of Hoyt \emph{et al.} \cite{Hoyt:00}.

\begin{table}[t]
\centering
\begin{tabular}{rrrr}
Metal & $-c(r=0)$ & $-\tilde{c}(k=0)$ &
$\displaystyle\frac{c(r=0)}{\tilde{c}(k=0)}$ \\
\hline
Cu & 60 & 47 & 1.3 \\
Ag & 51 & 53 & 1.0 \\
Au & 35 & 38 & 0.9 \\
Ni & 41 & 50 & 0.8
\end{tabular}
\caption{Direct correlation function $c(r=0)$ and its Fourier transform
$\tilde{c}(k=0)$ for four liquid transition metals. After
Ref.~\cite{Bernasconi:86}.}
\label{tab:Bernasconi}
\end{table}

A necessary condition for a liquid metal to be `hard-sphere'-like is for the
ratio $c(r=0)/\tilde{c}(k=0)$ to be near unity, as stressed by Bernasconi and
March \cite{Bernasconi:86}. This is well satisfied for Ag, Au, and Ni
(Tab.~\ref{tab:Bernasconi}). The polyvalent liquid metals Ga, Pb, and Sb are
obvious exceptions from the study of Bernasconi and March \cite{Bernasconi:86},
the ratios being respectively 0.2, 0.4, and 0.3.

Because of the results in Table~\ref{tab:Bernasconi}, we have been motivated to
return to the hard sphere model. However, we have transcended the
analytic model of Longuet-Higgins and Pople \cite{Longuet-Higgins:56} used
above, by invoking computer simulation results. Thus, Speedy \cite{Speedy:87}
has fitted molecular dynamics results for $D$ in the hard sphere model by the
approximate formula
\begin{equation}
D = D_0 \left( 1 - \frac{x}{1.09} \right) [1+x^4 (0.4 -0.83 x^4 )],
\label{eq:Speedy}
\end{equation}
where $x=\rho_N \sigma^3$ is proportional to the packing fraction $p_f$ introduced
above, while
\begin{equation}
D_0 = \frac{3\sigma}{8\rho_N} \left( \frac{\kB T}{\pi m} \right)^{1/2}
\end{equation}
is the infinite dilution value of $D$, which is known exactly from kinetic
theory.

\begin{figure}[t]
\centering
\includegraphics[height=0.9\columnwidth,angle=-90]{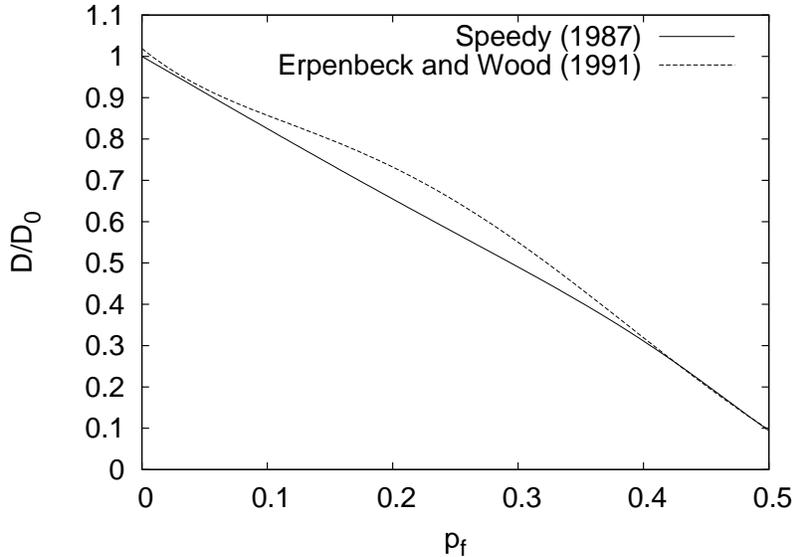}
\caption{Normalized self-diffusion coefficient $D/D_0$ as a function of packing
fraction $p_f$ within the hard sphere model, according to the Speedy
[Eq.~(\ref{eq:Speedy})] and Erpenbeck and Wood [Eq.~(\ref{eq:Wood})] results.}
\label{fig:selfdiff}
\end{figure}

\begin{figure}[t]
\centering
\includegraphics[height=0.9\columnwidth,angle=-90]{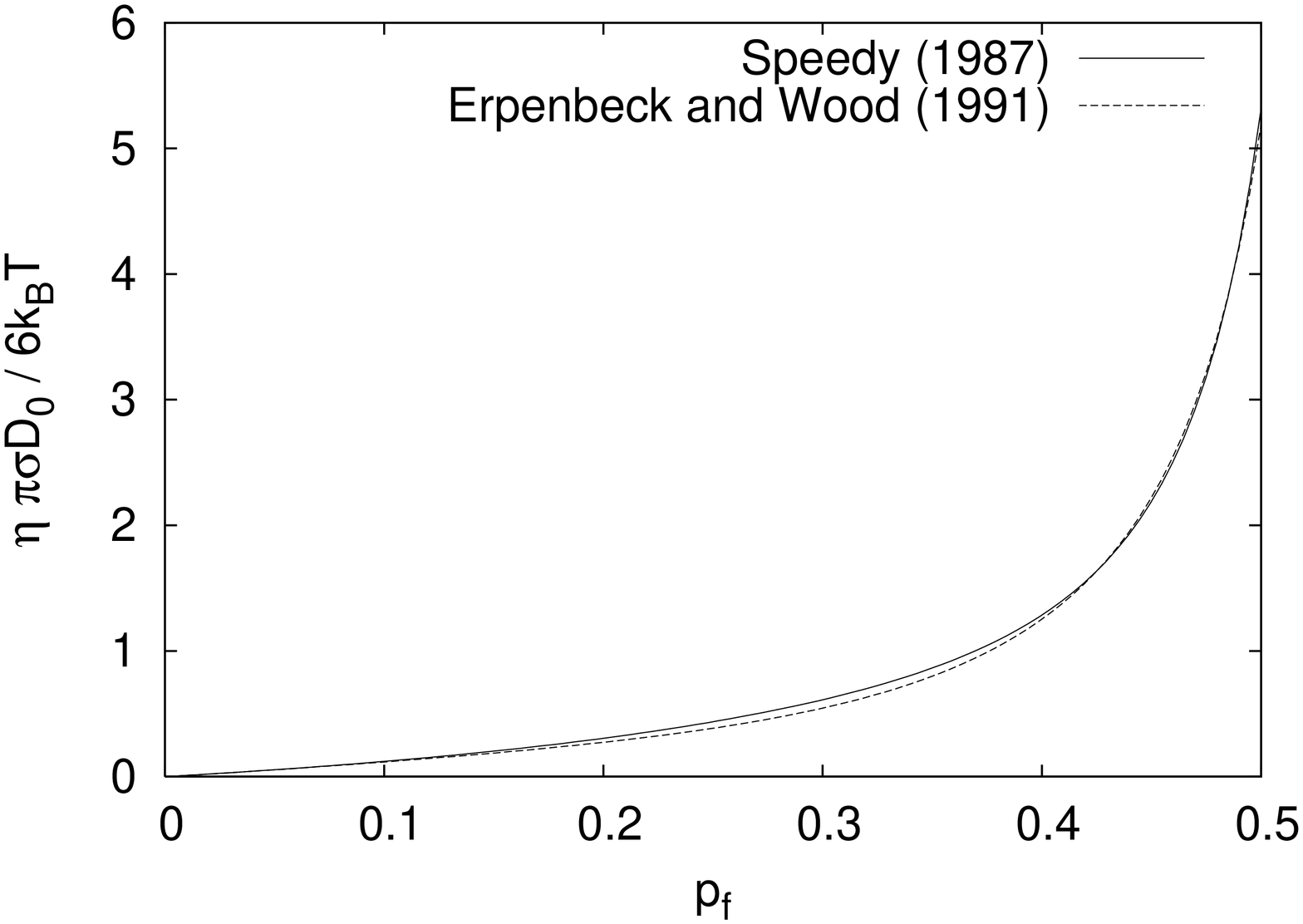}
\caption{Approximation to hard sphere viscosity $\eta_{\mathrm{HS}}$ using
computer results for $D/D_0$ in Fig.~\ref{fig:selfdiff}, plus
Eq.~(\ref{eq:Stokes}).}
\label{fig:Einstein}
\end{figure}

For comparison with Eq.~(\ref{eq:Speedy}) due to Speedy \cite{Speedy:87}, an
alternative expression for the density dependence of the self-diffusion for hard
spheres has been given by Erpenbeck and Wood \cite{Erpenbeck:91} as
\begin{equation}
D = D_E (1+0.0382 x + 3.18 x^2 -3.869 x^3 ),
\label{eq:Wood}
\end{equation}
where
\begin{equation}
D_E = \frac{1.01896 D_0}{g_{\mathrm{hs}} (\sigma^+ )} ,
\end{equation}
$g_{\mathrm{hs}} (\sigma^+ )$ denoting the hard sphere pair distribution
function at contact, referred to earlier in relation to the Longuet-Higgins and
Pople \cite{Longuet-Higgins:56} study. This is related to the packing fraction
$p_f$ by
\begin{equation}
g_{\mathrm{hs}} (\sigma^+ ) = \frac{1-p_f /2}{(1-p_f )^3} .
\end{equation}
Eqs.~(\ref{eq:Speedy}) and (\ref{eq:Wood}) are then compared in
Fig.~\ref{fig:selfdiff} as a function of packing fraction $p_f$. Since our main
focus here is the shear viscosity $\eta$, we have used the, of course
approximate, proposal of Longuet-Higgins and Pople \cite{Longuet-Higgins:56} for
the specific form of the Stokes-Einstein relation, \emph{viz.}
\begin{equation}
D\eta = \frac{1}{10} \sigma^2 \rho_N \kB T,
\label{eq:Stokes}
\end{equation}
to plot $\eta$ \emph{vs} $p_f$ in Fig.~\ref{fig:Einstein}.

\subsection{Scaled experimental data on liquid alkali metals}
\label{sec:alkali}

To conclude this section, it seemed of interest to refer to scaling properties
of the five liquid alkali metals Li, Na, K, Rb, and Cs \cite{Tankeshwar:98}. Why
the liquid alkali metals are especially important in the present context is
clear from the scaling exhibited in \cite{Tankeshwar:98} of the shear viscosity
$\eta(T)$ from available experiments. In fact the so-called fluidity $\eta^{-1}$
is the quantity plotted, but appropriately scaled by its value at the melting
temperature $T_{\mathrm{melt}}$, to distinguish from the minimum $\Tmin$ in the
plot of the ratio $\eta/s$ \emph{vs} $T$ (Fig.~\ref{fig:Kovtun}).

\begin{figure}[t]
\centering
\includegraphics[height=0.9\columnwidth,angle=-90]{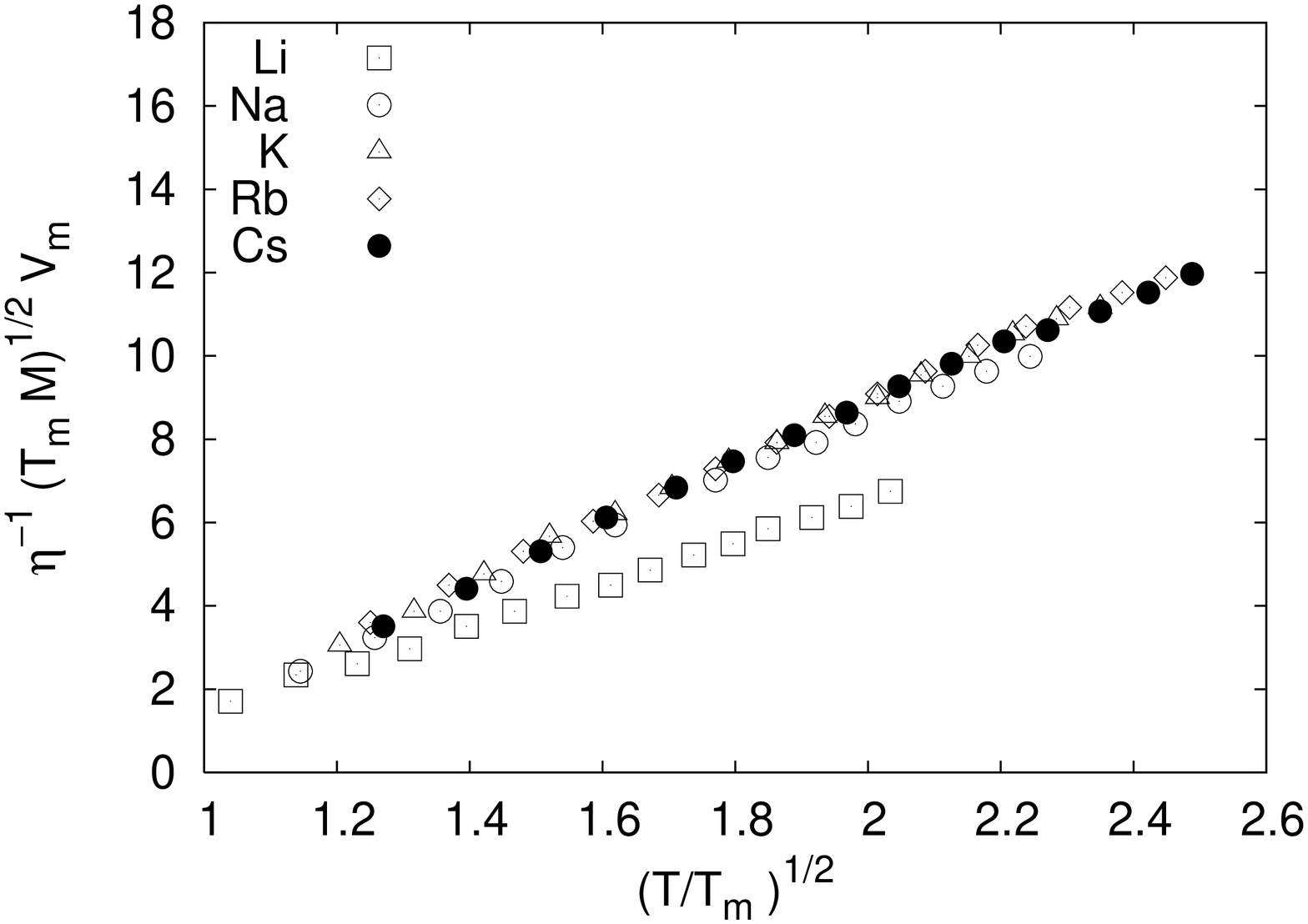}
\caption{Scaled inverse of shear viscosity $\eta(T)$ as a function of temperature $T$ for
liquid alkali metals (redrawn from Ref.~\cite{Tankeshwar:98}). Note that, apart
from the lightest metal Li, the other data fall on an `almost universal' curve.
Note that $T_m$ is the melting temperature.}
\label{fig:shear}
\end{figure}

We have redrawn Fig.~3 of Ref.~\cite{Tankeshwar:98} in Fig.~\ref{fig:shear} and
it is quite clear that, for the four heavier alkalis, Na, K, Rb, and Cs, the
scaling used collapses the data onto an `almost universal' curve. While there is
some shape similarity for the remaining (lightest) alkali metal Li, this has an
appreciably different magnitude, under scaling, from the other four. In
principle, one can estimate the volume density of entropy $s$ as a function of
$T$ from experimental structure factor data.

\begin{figure}[t]
\centering
\includegraphics[height=0.45\columnwidth,angle=-90]{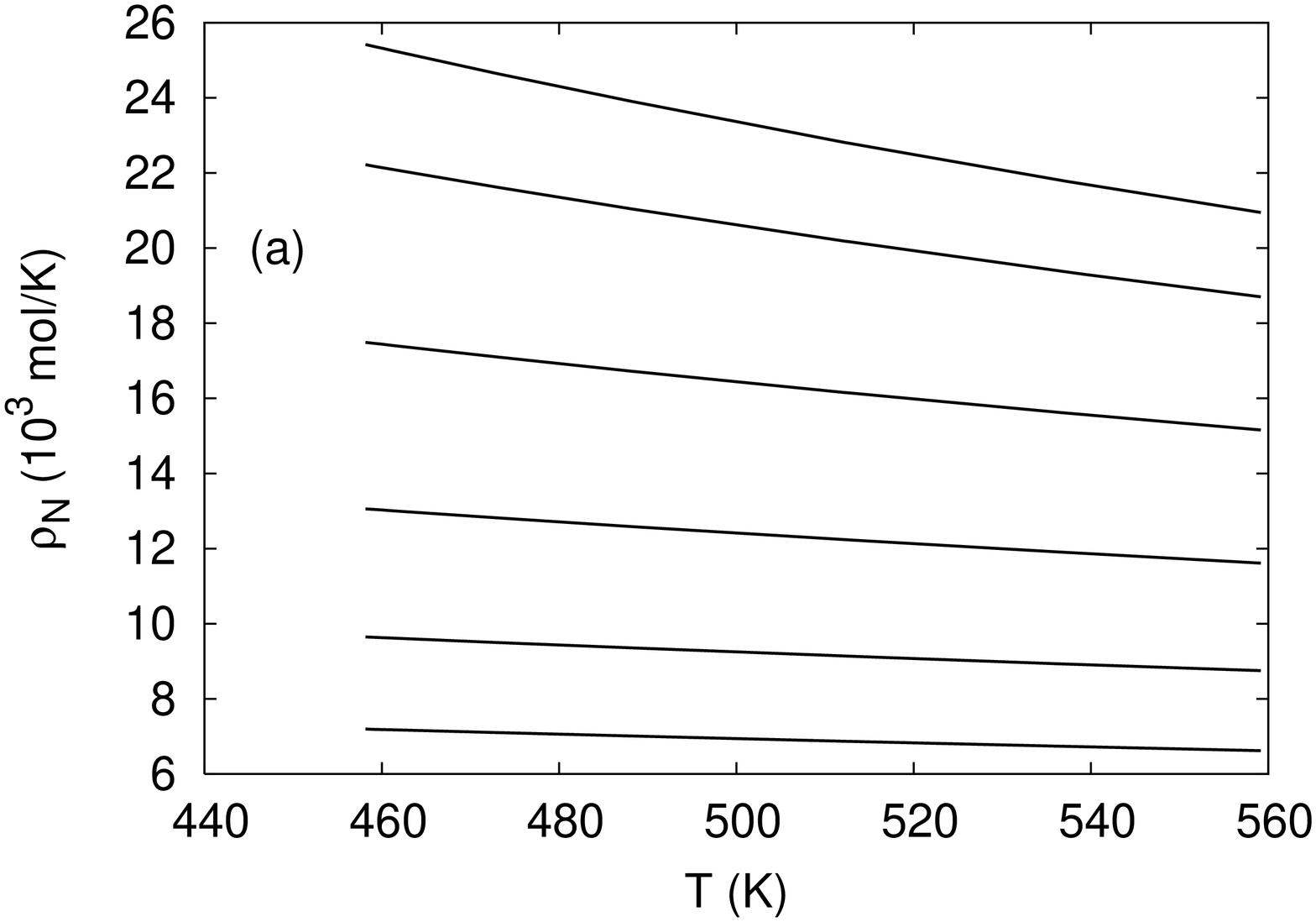}
\includegraphics[height=0.45\columnwidth,angle=-90]{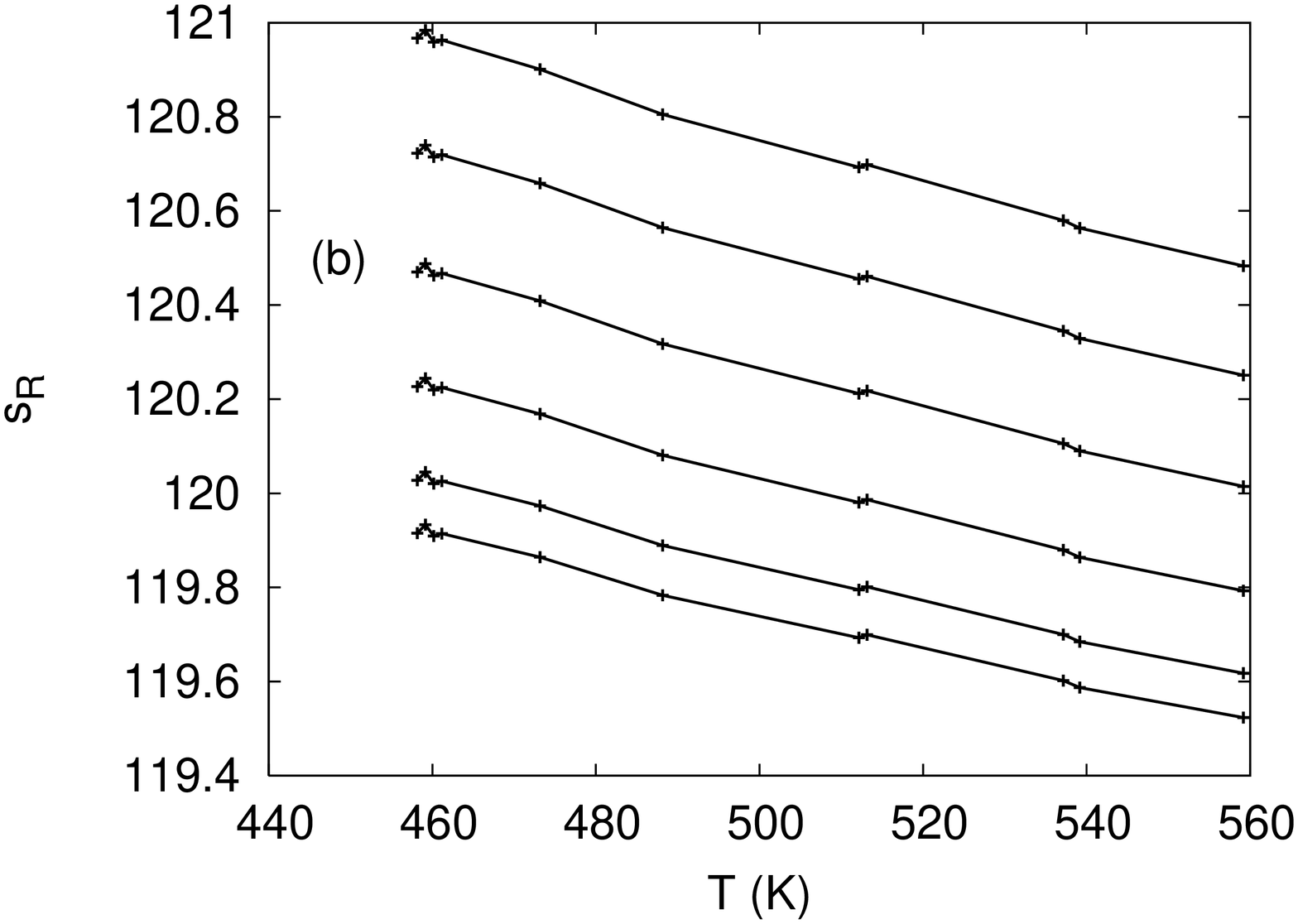}\\
\includegraphics[height=0.45\columnwidth,angle=-90]{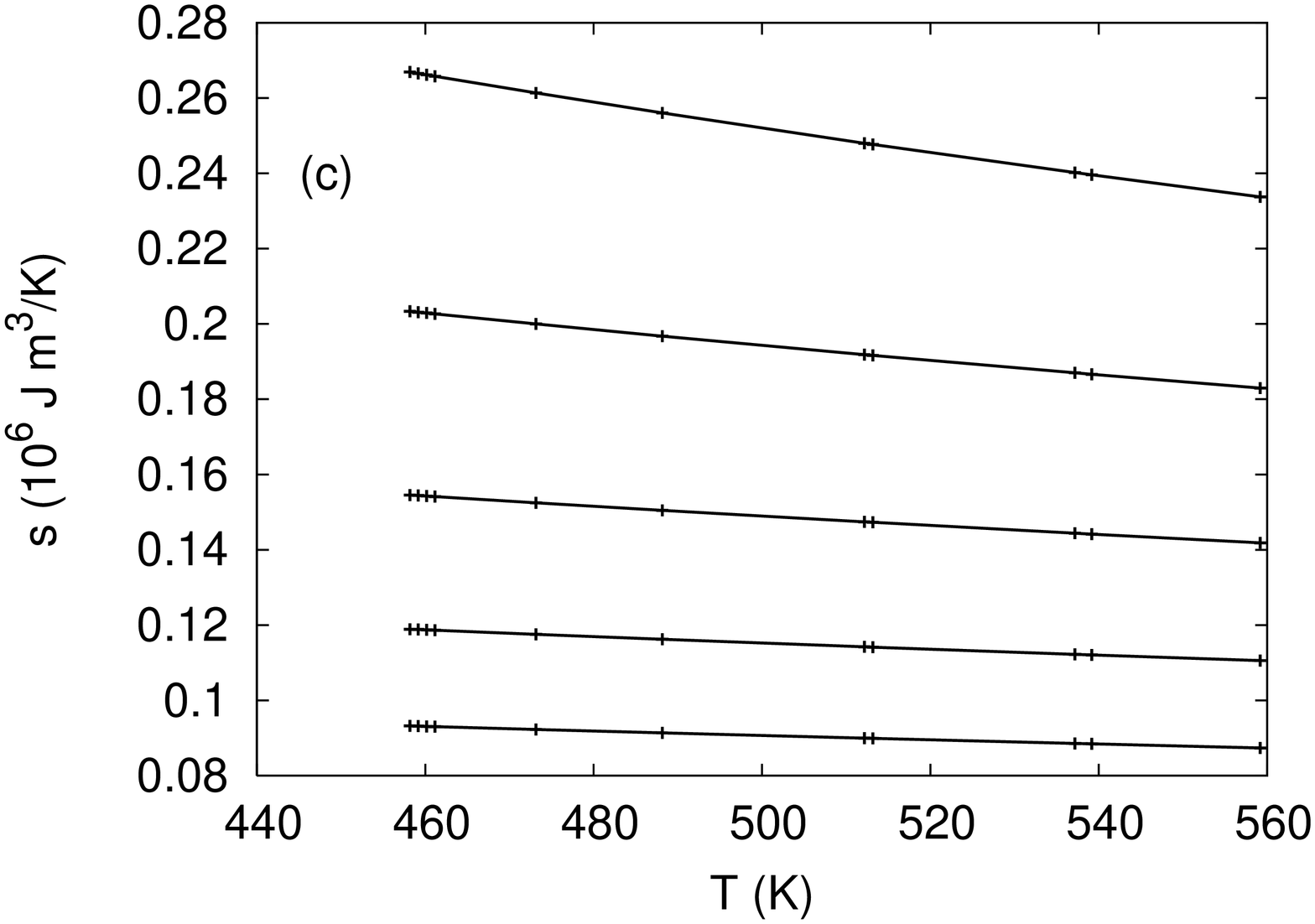}
\caption{(a) Number density $\rho_N$ for Li, employing Carnahan
equation of state, corrected using one-component plasma theory
\cite{Chapman:86}. (b) Rosenfeld entropy $s_{\mathrm{R}}$, as in
Eq.~(\ref{eq:etareds}), using data for viscosity $\eta$ for Li in \cite{Ban:62}.
(c) Volume density of entropy, using Eq.~(\ref{eq:S0}).
Pressure is 100~MPa, while $\sigma=1-5$~\AA.}
\label{fig:entropy}
\end{figure}

In fact, in a further paper \cite{Rosenfeld:99}, Rosenfeld writes an empirical
formula relating shear viscosity $\eta$ and scaled entropy $s_{\mathrm{R}}$, where
\begin{equation}
s_{\mathrm{R}}=-(S-S_0 )/N\kB ,
\label{eq:sR}
\end{equation}
$S_0$ referring to the ideal gas as in Eq.~(\ref{eq:tetrahedra}). Writing \cite{Rosenfeld:99}
\begin{equation}
\eta_{\mathrm{red}} = \eta \rho_N^{-2/3} / (M\kB T)^{1/2} ,
\end{equation}
where $\rho_N$ is the particle number density, while $M$ is the ionic mass,
Rosenfeld proposes the semiempirical formula
\begin{equation}
\eta_{\mathrm{red}} = 0.2 \exp (0.8 s_{\mathrm{R}}).
\label{eq:etared}
\end{equation}
Hence, it follows \cite{Rosenfeld:99} that
\begin{equation}
\frac{\eta}{s} = 0.2 \rho_N^{2/3} (M\kB T)^{1/2} \frac{\exp(0.8 s_{\mathrm{R}})}{s} .
\label{eq:etareds}
\end{equation}
We have employed the experimental viscosity data in Ref.~\cite{Ban:62} for
liquid Li (see also Fig.~\ref{fig:shear}), to extract the Rosenfeld scaled
excess entropy $s_{\mathrm{R}}$ from Eq.~(\ref{eq:etareds}), when combined with
the approximate equation of state of Chapman and March \cite{Chapman:86}. This
reads
\begin{equation}
\frac{P}{\rho_N \kB T} = \frac{P_{\mathrm{HS}}}{\rho_N \kB T} -\alpha_n \Gamma
\end{equation}
where $\Gamma$ is the coupling parameter of the so-called one component plasma
(see \cite{March:84b}) defined by
\begin{equation}
\Gamma = \frac{e^2}{a\kB T} ,
\end{equation}
where $\rho_N^{-1} =4\pi a^3 /3$. Fig.~\ref{fig:entropy}b shows schematic plots
of $\rho_N$ \emph{vs} $T$, which it would, of course, be of considerable
interest to have quantitative experimental results in the future at the
thermodynamic conditions in Fig.~\ref{fig:shear} and in the experimental
study of $\eta(T)$ in Ban \emph{et al.} \cite{Ban:62}. Fig.~\ref{fig:entropy}c
finally shows the schematic form of the volume density of entropy $s(T)$
obtained from Eqs.~(\ref{eq:Rosenfeld}) and (\ref{eq:S0}).

\section{Discussion and future directions}
\label{sec:summary}

Motivated by the proposal of Kovtun \emph{et al.} \cite{Kovtun:05} of a lower
bound for the ratio of shear viscosity $\eta$ to volume density of entropy $s$,
we have studied this ratio for some fifteen dense liquids.  These include liquid
metals, five liquid alkalis, plus some six liquid transition metals, for the
former of which the shear viscosity $\eta(T)$ has been determined experimentally
over a range of temperature $T$. Using the scaling of these data in
\cite{Tankeshwar:98}, Figure~\ref{fig:shear} has been constructed by making use
of the semiempirical Eq.~(\ref{eq:etared}) given by Rosenfeld
\cite{Rosenfeld:77,Rosenfeld:99}. Into this equation, the experimental data for
$\eta(T)$ has first been used to predict the volume density of entropy $s(T)$
plotted in Fig.~\ref{fig:entropy}. It will be of significance for the future to
refine these data using X-ray or neutron diffraction measurements to estimate
semiempirically the pair correlation function $g(r)$. This is then sufficient to
determine the entropy $s$, following, for instance, the procedure of Dzugotov
\cite{Dzugotov:96}.

The most quantitative results in the present study are presented in
Fig.~\ref{fig:Kovtun}. In part (a), we have scaled data for the four insulating
liquids shown by plotting the ratio $\eta/s$ \emph{vs} $T/T_{\mathrm{min}}$,
where $T_{\mathrm{min}}$ is the temperature at the minimum of $\eta/s$ in each
case. It is seen that there is considerable shape similarity for the four
liquids, in spite of the range of isobaric conditions involved.
Fig.~\ref{fig:Kovtun}b is presented, again from experiment, to give insight into
the origin of the minima in $\eta/s$. Over the range plotted in
Fig.~\ref{fig:Kovtun}b, $\eta(T)$ is monotonically decreasing with increasing
temperature, and the minimum in $\eta/s$ therefore arises from the turning point
in the volume density of entropy $s(T)$, at temperature $T_{\mathrm{max}}$.
Fig.~\ref{fig:Kovtun}c shows that both $T_{\mathrm{min}}$ and $T_{\mathrm{max}}$
are of the order of the critical temperature $T_c$ for N$_2$, H$_2$O, and
NH$_3$.

The remainder of the article begins a study of the ten liquid metals already
mentioned. Because of the incomplete availability of (i) experiment and (ii)
computer simulation data, we have had recourse to models. These are (i) the well
known HS model, set out in Section~\ref{sec:hs}, where both approximate analytic
theory and computer simulation data is now available (see especially
Fig.~\ref{fig:selfdiff} for the diffusion coefficient $D$). This has been
combined with the (of course approximate) analytic form of the Stokes-Einstein
Eq.~(\ref{eq:Stokes}) to construct Fig.~\ref{fig:Einstein}. It would, of course,
be valuable to have results for the shear viscosity $\eta$ paralleling those
displayed in Fig.~\ref{fig:selfdiff} for $D$, to test the HS predictions of
Fig.~\ref{fig:Einstein}, and if necessary to refine them.

We may mention also at this point the generalization Eq.~(\ref{eq:tetrahedra})
of the HS equation of state (\ref{eq:Carnahan}). It has been suggested
\cite{Kolafa:95} that Eq.~(\ref{eq:tetrahedra}) may have some relevance to water
with $\alpha\simeq 2$, which is close to the numerical value 1.7 employed to
construct Fig.~\ref{fig:rhoN}, relating to experiments on water and ammonia.
While on the HS model, the `softening' of the core to inverse power repulsive
potentials is referred in Appendix~\ref{app:invpot} (see also especially
Fig.~\ref{fig:invrep}).

For the ten or so liquid metals considered, and beginning with the six
transition elements studied by Hoyt \emph{et al.} \cite{Hoyt:00} using the
embedded atom model, Table~\ref{tab:Bernasconi} shows that a necessary, though
not sufficient, condition for HS behavior (that the final column is equal to
unity) is well obeyed for two, Ag and Au, and is approximately valid for Cu and
Ni. So in the future, it will be interesting to bring computer simulation and
experiment into contact with the HS model, and especially with the predictions
for the shear viscosity $\eta$ in Fig.~\ref{fig:Einstein}. Finally, for the five
liquid alkalis, Fig.~\ref{fig:shear} summarizes scaling properties of the shear
viscosity. However, experimental results on the density-temperature relation
under the thermodynamic conditions appropriate to Fig.~\ref{fig:shear} and to
the isotope experiments of Ban \emph{et al.} \cite{Ban:62} would be most helpful
in the present context, as, next to of course He (see Fig.~\ref{fig:Kovtun}), Li
is the most likely candidate for quantal effects, and thus for a possible lower
bound to compare with the prediction of Kovtun \emph{et al.} \cite{Kovtun:05}.

In a related vein, we note that the work of Tosi \cite{Tosi:92} on electron
plasma theory has led to the conclusion that the itinerant electrons in the
alkali metals contribute less than 10\% to the shear viscosity. As a proposal
for future work, though somewhat esoteric, it would be of interest to study in
its own right the homogeneous electron liquid \cite{Giuliani:05}.

As for liquid Li, we believe it would be of considerable interest to use a
density-dependent pair potential, such as that derived by Parrot and March
\cite{Perrot:90} from first principles electron theory, to calculate the excess
entropy approximately from the resulting pair distribution function $g(r)$, as
set out in \cite{Dzugotov:96}. Then, adding the ideal entropy in
Eq.~(\ref{eq:S0}), one could plot the volume density of entropy, say, along an
isobar. This could then be compared for any shape similarity with the three
insulating liquids treated in Fig.~\ref{fig:Kovtun}b.   

As this study was nearing completion, we became aware of the highly relevant
work of Sch\"afer \cite{Schaefer:07}. This author studied the ratio $\eta/s$
for trapped fermions in the unitarity limit. The result of this investigation
was the prediction that $\eta/s$ is roughly $1/2$ in units of $\hbar/\kB$. This
example prompts us to conclude this article by summarizing briefly the uncertainty
principle argument pointed out by Kovtun \emph{et al.} \cite{Kovtun:05}. This
consists of three parts: (i) $\eta$ for a plasma is proportional to
$\epsilon\tau$, where $\epsilon$ is the energy density while $\tau$ is the
typical mean free time of a quasiparticle, and (ii) the entropy density is
proportional to the density of quasiparticles, say $n$, \emph{i.e.} $s\sim \kB
n$. Thus the desired ratio $\eta/s \sim \kB^{-1} \epsilon\tau/n$. The third part
of the argument \cite{Kovtun:05} invokes, as mentioned above, the uncertainty
principle in the form $(\epsilon/h)\tau > \hbar$, without which the concept of
quasiparticle lacks meaning. This gives $\eta/s \gtrsim \hbar/\kB$, without,
however, the factor $1/4\pi$ in \cite{Kovtun:05}.

\section*{Acknowledgements}

NHM began his contribution to this study at the Abdus Salam International Centre
for Theoretical Physics. It is a pleasure to thank Professor V. E. Kravtsov for
generous hospitality. The work of NHM was brought to fruition during a stay at
the University of Catania. NHM was partially supported by the University of
Antwerp through the BOF-NOI.

\appendix

\section{Scaled diffusion coefficient in terms of excess entropy $S_E$ in the
hard sphere model: Relation to the work of Kovtun \emph{et al.}}
\label{app:D}

In Fig.~\ref{fig:selfdiff} we compare the computer simulation results of Speedy
\cite{Speedy:87} with those of Erpenbeck and Wood \cite{Erpenbeck:91}.
Evidently, with refinements to available computer data, the two curves in
Fig.~\ref{fig:selfdiff} will converge to a single curve which we represent by
writing the scaled diffusion coefficient $D/D_0$ in the functional form
\begin{equation}
\frac{D}{D_0} = f(p_f ).
\label{eq:DD0}
\end{equation}
But for this HS model, we know the excess entropy $S_E$ in the Percus-Yevick
approximation as is given in Eq.~(\ref{eq:excess}). From Fig.~\ref{fig:selfdiff}
representing an approximation to the formally exact Eq.~(\ref{eq:DD0}), we can
conclude that a unique inversion exists, to yield
\begin{equation}
p_f = g(D/D_0 ).
\label{eq:invDD0}
\end{equation}
Inserting Eq.~(\ref{eq:invDD0}) into Eq.~(\ref{eq:excess}), we therefore
conclude that there exists the functional relation
\begin{equation}
\frac{S_E}{\kB} = h (D/D_0 )
\label{eq:EDD0}
\end{equation}
in the HS model.

If we combine this exact formal result, Eq.~(\ref{eq:EDD0}) with the (now
approximate) Stokes-Einstein relation proposed by Longuet-Higgins and Pople
\cite{Longuet-Higgins:56} for the HS model, Eq.~(\ref{eq:Stokes}), and use $p_f
= \pi \rho_N \sigma^3 /6$ to find
\begin{equation}
D\eta = \frac{3}{5\pi} \frac{p_f}{\sigma} \kB T ,
\label{eq:Deta1}
\end{equation}
we find from Eqs.~(\ref{eq:DD0}) and (\ref{eq:Deta1}) that
\begin{equation}
\eta = \left( \frac{D_0}{D} \right) \left( \frac{3}{5\pi} \frac{p_f}{\sigma}
\frac{\kB T}{D_0} \right).
\label{eq:Deta2}
\end{equation}
But with the plausible assumption that Eq.~(\ref{eq:EDD0}) has the unique
inversion
\begin{equation}
\frac{D}{D_0} = j \left( \frac{S_E}{\kB} \right) ,
\end{equation}
we find by combining this equation with Eq.~(\ref{eq:Deta2}) that
\begin{equation}
\frac{\eta\sigma D_0}{\kB T} = \ell \left( \frac{S_E}{\kB} \right).
\label{eq:Deta3}
\end{equation}
We propose that Eq.~(\ref{eq:Deta3}) demonstrates the functional form of $\eta$
in terms of the dimensionless excess entropy $S_E /\kB$ in the HS model.

But when we turn to consider the ratio $\eta/s$ as in the lower bound proposed
by Kovtun \emph{et al.} \cite{Kovtun:05}, we note that the total entropy $S$ is
given by Eq.~(\ref{eq:sR}). While $S_E /\kB$ in Eq.~(\ref{eq:EDD0}) is solely a
function of the packing factor $p_f$, $S_0$ involves also, following Landau and
Lifshitz \cite{Landau:80}, the thermal de~Broglie wavelength $\lambda =
(2\pi\hbar^2 / M\kB T)^{1/2}$, providing a natural length unit for the volume
per particle $\rho_N^{-1}$, and hence the volume entropy $s=\rho_N S$ depends on
$\sigma$, $p_f$, and $\lambda$, as does the free energy $F$ since the ideal gas
result $F_0$ reads
\cite{Landau:80}
\begin{equation}
F_0 = - N\kB T [1-\log (\rho_N \lambda^3 )] 
\end{equation}
In the isobaric plots of Fig.~\ref{fig:Kovtun}, $\rho_N = \rho_N (T)$, and one
has that $\eta/s \equiv n(T)$.

\section{Results for inverse power repulsive potentials}
\label{app:invpot}

In the main text, together with experiments and their interpretation especially
on the molecular insulating liquids nitrogen, water, and ammonia, we have given prominence
to the hard sphere model. Here, we generalize this model to the case of inverse
power repulsive potentials having the form
\begin{equation}
\phi(r) = \epsilon \left( \frac{\sigma}{r} \right)^n .
\label{eq:pairn}
\end{equation}
The merit of this model is that, for any such inverse power potential,
thermodynamic properties are readily accessible since only a single isotherm,
isochoric, or isobar needs to be known. Then all other properties can be found,
as discussed by Hoover and Ross \cite{Hoover:71}.

Quite specifically, for pair-wise additive potentials having the form
(\ref{eq:pairn}), (now dimensionless) thermodynamic properties depend only on
the one density-demperature variable, $y$ say:
\begin{equation}
y = \rho\left( \frac{\epsilon}{\kB T} \right)^{3/n} ,
\end{equation}
where $\rho$ is proportional to the packing fraction $p_f$ and is defined as
$\rho =N\sigma^3 /V$. 

\begin{figure}[t]
\centering
\includegraphics[height=0.9\columnwidth,angle=-90]{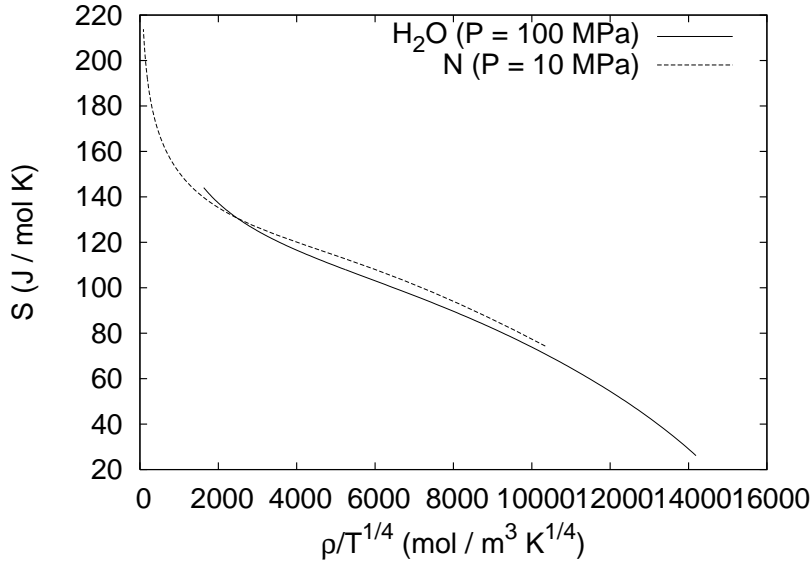}
\caption{Entropy for water and nitrogen at the given pressures (data source as
in Kovtun \emph{et al.} \cite{Kovtun:05}), as a function of $\rho/T^{3/n}$, for
$n=12$. Data for water have been shifted vertically for clarity.}
\label{fig:invrep}
\end{figure}

We have studied the data on nitrogen and water presented in
Sec.~\ref{sec:liquids} of the main text, and for $n=12$ we find remarkable shape
similarity (Fig.~\ref{fig:invrep}).

Because of transport coefficients being another focal point here, we finish this
Appendix by referring to the interesting study of the diffusion coefficient
$D(n)$ for the potential (\ref{eq:pairn}) by Heyes and Powles \cite{Heyes:98}. For
a packing fraction of 0.044, these authors studied how $D(n)$ approached the
hard sphere limit $D(\infty)$ and found that their data could be well fitted by
the form, with $D$ now in dimensionless form,
\begin{equation}
D(n)-D(\infty) = \mathrm{const} \cdot n^{-1.71} ,
\end{equation}
indicating a rather slow approach to the hard sphere result.

\bibliographystyle{mprsty}
\bibliography{a,b,c,d,e,f,g,h,i,j,k,l,m,n,o,p,q,r,s,t,u,v,w,x,y,z,zzproceedings,Angilella}

\end{document}